\documentclass[aps,pre,reprint,twocolumn,showpacs,amsmath,amssymb,%
superscriptaddress]
{revtex4-1}
\usepackage{graphicx}
\usepackage{color}
\begin{document}
\title{Dissecting a Small Artificial Neural Network}
\author{Xiguang Yang}
\affiliation{Soft Matter Systems Research Group, Center for Simulational
Physics, Department of Physics and Astronomy, The University of Georgia, 
Athens, GA 30602, USA}
\author{Krish Arora}
\affiliation{Soft Matter Systems Research Group, Center for Simulational
Physics, Department of Physics and Astronomy, The University of Georgia, 
Athens, GA 30602, USA}
\affiliation{College of Computing, Georgia Tech, Atlanta, GA 30332, USA}
\author{Michael Bachmann}
\email{bachmann@smsyslab.org; https://www.smsyslab.org}
\affiliation{Soft Matter Systems Research Group, Center for Simulational
Physics, Department of Physics and Astronomy, The University of Georgia, 
Athens, GA 30602, USA}
\begin{abstract}
We investigate the loss landscape and backpropagation dynamics of convergence 
for the simplest possible artificial neural network representing the logical 
exclusive-OR (XOR) gate. Cross-sections of the loss landscape in the 
nine-dimensional parameter space are found to exhibit distinct features, 
which help understand why backpropagation efficiently achieves convergence 
toward zero loss, whereas values of weights and biases keep drifting. 
Differences in shapes of cross-sections obtained by nonrandomized and 
randomized batches are discussed.
In reference to statistical physics we 
introduce the microcanonical entropy as a unique quantity that allows to 
characterize the phase behavior of the network. Learning in 
neural networks can thus be thought of as an annealing process that 
experiences the analogue of 
phase transitions known from thermodynamic systems. It also reveals how the 
loss landscape simplifies as more hidden neurons are added to the network, 
eliminating entropic barriers caused by finite-size effects.
\end{abstract}
\maketitle 
\section{Introduction}
For more than half a century, it has been believed that an artificial 
intelligence can be developed to simulate the astonishing 
cognition, reasoning, and memory capabilities of the human brain. The 
network of a large number of neural cells connected through synapses 
apparently enables complex thinking and problem solving by processing input 
and generating distinct output signals. Cascades of electric activation 
pulses from neuron to neuron mediate the learning and memorizing of 
signal patterns that accomplish the process of decision making.

In simple artificial neural networks, the electric activations signals are 
replaced by a large matrix of weights and biases that serve as input 
variables for a nonlinear activation function at each neuron. Basic networks 
are composed of a layer of input neurons, potentially multiple layers of 
hidden neurons to create neural capacity, and a layer of output neurons. 
Learning is achieved by an efficient recursive method called 
backpropagation~\cite{kelley1,bryson1,hinton1,werbos1}. 
Besides explicit mathematical calculation and computational logical 
algorithms, artificial neural networks can be considered an additional class 
of problem-solving strategies~\cite{russell1}.

Early attempts to employ artificial neural networks date back to the late 
1950s. The first such network was the perceptron, which consisted of
input and output neurons only~\cite{rosenblatt1}. Although the potential of 
neural 
networks was recognized early, the perceptron was too 
simple~\cite{minsky1,hinton2}. The fact that 
it could not be used to represent the logical XOR gate is 
believed to have caused the ``first AI winter'' at the end of the 
1960s. Even 
though significantly improved computational and image processing equipment 
and algorithmic advances lead to a renewed interest through the 1980s, actual 
useful applications did not develop beyond so-called expert machines capable 
of only solving very specific problems. Only when the global internet made 
computational networking possible at a large scale, leading to the creation 
and storage of vast amount of data, the pressing need of processing these 
data forced a breakthrough in the further development of artificial 
intelligence in the 2010s. The large networks necessary to process these 
data are trained in a process called deep learning, which, 
however, is still 
based on backpropagation. A large variety of neural network architectures 
have been proposed to accelerate the learning 
process~\cite{russell1,herberg1}. 

Studies of the Hopfield model~\cite{hopfield1} were among the first to 
show the deep relationship of complexity in neural networks and 
disordered magnetic systems. This triggered substantial interest in the 
statistical physics community and culminated 
in a vast literature on the subject in past decades. Most studies focused on 
understanding the many factors that can impede the performance of the 
optimization or learning process, though. Limiting factors are not only 
the 
architecture of the network and the optimization strategy 
themselves~\cite{herberg1,liaoa1}, but also 
the choice of initial conditions, input data formats, and batching.

Lesser effort has been dedicated thus far to the 
understanding of the geometry of the underlying state space 
and features of the energy or loss landscape and its generalization by means 
of the density of states. The latter notoriously defies analytic treatment 
and, therefore, only recently found increasing attention in computational 
studies of phase transitions in finite systems. These approaches have been 
proven useful in applications to problems in statistical 
physics~\cite{gori1,dicairano1}.

To further advance our understanding of the complex nature of neural 
networks, we also establish the relationship 
between the two driving forces of phase transitions, energy and entropy, and 
determine the microcanonical entropy. Inflection points in the entropy curve 
have been found to be reliable indicators for phase transitions~\cite{qb1}, 
even in finite systems such as biomolecules~\cite{ab1}. 

In order to make artificial intelligence a safe assistant of humanity, it is 
necessary to deeply understand how artificial neural networks learn and work. 
Given that modern complex networks often comprise billions of neurons, it is 
impossible to track the learning process and function in all detail. 
Therefore, in this paper, we take the opposite approach and study one of the 
smallest possible meaningful networks that is capable of solving the XOR 
problem. Besides 
the two input neurons and the single output neuron, only one additional layer 
of two hidden neurons is needed. Effectively, there are nine parameters to 
be 
tuned. The simplicity enables us to perform an excessive number of tests and 
measurements incomprehensible for large networks. 

Yet, the learning 
procedures and therefore the convergence processes are similar and should 
give 
us general insights into the actual working of an artificial neural network.
Particular emphasis is dedicated to features of the loss landscape embedded 
into the high-dimensional parameter space. Visualizing cross-sections 
through the landscape yields valuable insights that may help improve the 
performance of the parameter optimization approach. Loss landscapes are 
comparable to effective potential-energy and free-energy landscapes in the 
spaces of relevant degrees of freedom or order parameters in physical 
systems~\cite{mb1}. Consequently, different approaches to the 
parameterization and study of the complexity of loss landscapes have been 
followed~\cite{wales1,wales2,goldstein1,grohs1}, with the purpose of better 
understanding the influence of the landscape shape upon the dynamics of 
optimization processes. In this study, we pursue a more straightforward 
approach and investigate features of the loss landscape 
impacting the dynamics of backpropagation convergence, which turns 
out to be surprisingly complex.

The paper is organized as follows. In Sect.~\ref{sec:ann}, we introduce the 
neural network and XOR logic studied and review the optimization by 
backpropagation. The results we obtained from the analysis of the 
convergence dynamics, the features of the loss landscape, as well as the 
discussion of the microcanonical entropies for networks with different 
numbers of hidden neurons, are presented in Sect.~\ref{sec:sol}. The summary 
in 
Sect.~\ref{sec:sum} 
concludes the paper. 
\section{XOR Neural Network Model and Optimization}
\label{sec:ann}
In this study, we employ a simple artificial neural network to solve the XOR 
problem. The binary inputs $x_1$ and $x_2$ and the output $y$ for all four 
cases are listed in Table~\ref{tab:xor}. It is straightforward to modify this 
basic gate for any other logical behavior, or to add complexity, for example 
by simultaneously considering multiple outputs for different logical 
operations. However, in order to be able to thoroughly analyze the 
optimization of the network in detail, we keep it as simple as possible.
\begin{table}[ht]
\caption{\label{tab:xor}Truth table of the XOR logic.}
\begin{tabular}{|cc|c|}
\hline
\hspace*{3mm}$x_1$\hspace*{3mm} & \hspace*{3mm} $x_2$ \hspace*{3mm}  & 
\hspace*{3mm}$y$ \hspace*{3mm}\\ \hline
$0$   & $0$   & $0$ \\
$0$   & $1$   & $1$ \\
$1$   & $0$   & $1$ \\
$1$   & $1$   & $0$ \\ \hline
\end{tabular}
\end{table}

\begin{figure}
\centerline{\includegraphics[width=6.0cm]{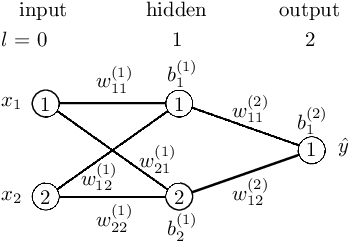}}
\caption{\label{fig:ann}%
Parametrization of the simplest artificial neural network used in this 
study 
with only $n_\mathrm{h}=2$ neurons in the hidden layer.}
\end{figure}

Since the perceptron, even with a single hidden neuron added, is known to 
lack capacity to memorize the four cases~\cite{hinton2}, we 
add more neurons to the hidden layer. Throughout the paper, we denote the 
number of hidden neurons by $n_\mathrm{h}$. The simplest network  we study 
here now consists of two 
neurons in the input layer, two neurons in the single hidden layer 
($n_\mathrm{h}=2$), and the 
output neuron (see Fig.~\ref{fig:ann}). In total, the fully connected 
network 
has six links with a weight parameter attached to each. Hidden and output 
neurons are assigned a bias parameter. Thus, the optimization problem to find 
an approximate solution that mimics the truth table in Table~\ref{tab:xor} 
is nine-dimensional. For comparison, larger networks with up to 
$n_\mathrm{h}=18$ hidden neurons have been studied as well.

We denote the weight associated with the link that connects the $j$th neuron 
in the $(l-1)$th layer with the $i$th neuron in layer $l$ by $w_{ij}^{(l)}$, 
which can be considered an element of the weight matrix $\mathbf{W}^{(l)}$. 
For the input layer, we set $l=0$. In our simple network, $l=1$ is the hidden 
layer and $l=2$ the output layer. The bias assigned to the $i$th neuron in 
layer $l$ is symbolized by $b_i^{(l)}$. It is useful to also introduce the 
auxiliary vector
\begin{equation}
\mathbf{z}^{(l)}=\mathbf{W}^{{(l)}}\mathbf{a}^{(l-1)}+\mathbf{b}^{(l)}
\end{equation}
with elements
\begin{equation}
z_i^{(l)}= \sum_{j=1}^{n_i^{(l-1)}} w_{ij}^{(l)}a_j^{(l-1)}+b_i^{(l)},
\label{eq:z}
\end{equation}
where $n_i^{(l-1)}$ is the number of neurons in the previous layer linked to 
neuron $j$ in layer $l$. The variable $a_i^{(l)}$ represents the 
activation value of the $j$th neuron in layer $l$. To determine the 
activation, we employ the sigmoid function 
\begin{equation}
a_i^{(l)}=\sigma\left(z_i^{(l)}\right), \quad l>0,
\label{eq:activ}
\end{equation} 
where $\sigma(z)=1/(1+e^{-z})$. For the input neurons, $a_i^{(0)}\equiv x_i$. 
The sole purpose of introducing the activation function is to break the 
linearity in Eq.~(\ref{eq:z}). Hence, other choices than $\sigma(z)$ are also 
popular, foremost the simpler ReLU (rectified linear unit) function, which is 
mostly used in very large networks for its computational efficiency. However, 
for our purpose, the sigmoid function is ideal, because the range of output 
values in the interval $(0,1)$ suits binary problems well and, more 
importantly, the gradients to be calculated in the optimization process are 
well-behaved and easy to control, at least in small networks like ours. As it 
turns out, the sigmoid approach is more robust and requires far less 
fine-tuning of control parameters such as the 
learning rate in the optimization process than ReLU.

Since the optimal values of the weights and biases are not known in the 
beginning, we set them to random values initially. Then, for a given case
from Table~\ref{tab:xor} the activations for each neuron are calculated in 
the forward pass according to Eq.~(\ref{eq:activ}) from the input to the 
output side of the network. According 
to our setup, which does not contain a data post-processor at the output, the 
calculated output is simply $\hat{y}=a_1^{(2)}$. 

Consequently, one learning 
step 
at given values for weights and biases contains $M$ forward passes yielding 
outputs $\hat{y}_m$, $m=1,\ldots,M$.  
For our problem, a full batch consists of a set of $M=4$ cases as this is 
the dimension of the input data set given in Table~\ref{tab:xor}. The batch 
can 
either be composed by the complete, nonrandomized sequence of all cases 
listed in the truth table, or the four cases are randomly chosen from the 
list (therefore, a case may occur more than once in a randomized batch). Both 
approaches were compared in our study. As 
the total number of cases is very small, the introduction of 
mini-batches is not useful here. 

As a measure for the overall deviation of the calculated outputs from the 
true ones in the batch, we introduce the total loss function 
\begin{equation}
\label{eq:loss}
L=\frac{1}{M}\sum\limits_{m=1}^M {\cal L}_m,
\end{equation}
where
\begin{equation}
{\cal L}_m=(\hat{y}_m-y_m)^2
\end{equation}
is the contribution to the loss from the $m$th case in the batch. The network 
performs as desired if the total loss is sufficiently close to zero.

As the complexity of neural networks prevents any analytic approach at 
finding the optimal weight and bias values, it is common to use a recursive 
approximation scheme called 
backpropagation~\cite{kelley1,bryson1,hinton1,werbos1}. Given that $L$ 
inherently 
depends on all parameters, any deviation of $L$ from zero can be traced back 
to deviations of weights and biases from their optimal values. Hence, 
following the negative gradient of $L$ in the space of these variables 
and adjusting their values accordingly should help reduce the loss in 
subsequent learning steps. In order to control the rate of change, it is 
useful to introduce the learning rate $\eta$. If we use $\mathbf{v}$ as a 
generic vector, whose components $v_k$ are all weights or biases in the 
nine-dimensional parameter space represented by our network, then 
backpropagation updates the vector $\mathbf{v}$ from recursion (or epoch) 
$\tau$ to $\tau+1$ according to
\begin{equation}
\mathbf{v}^{(\tau+1)}=\mathbf{v}^{(\tau)}-\eta\frac{1}{M}\sum\limits_{m=1}^M
\nabla_\mathbf{v} {\cal L}_m\left(\mathbf{v}^{(\tau)}\right),
\end{equation}
where over the gradients, obtained from the individual cases in the batch, is 
averaged.
\begin{table*}[hbt]
\caption{\label{tab:sol}Examples of sets of optimal parameter values for 
the XOR network and sigmoid activation.}
\begin{tabular}{|c|c|c|c|c|c|c|c|c|c|}
\hline
\mbox{} & 
$\left[w_{11}^{(1)}\right]_\mathrm{opt}$ & 
$\left[w_{12}^{(1)}\right]_\mathrm{opt}$ & 
$\left[w_{21}^{(1)}\right]_\mathrm{opt}$ & 
$\left[w_{22}^{(1)}\right]_\mathrm{opt}$ &
$\left[b_1^{(1)}\right]_\mathrm{opt}$  & 
$\left[b_2^{(1)}\right]_\mathrm{opt}$  &
$\left[w_{11}^{(2)}\right]_\mathrm{opt}$ & 
$\left[w_{12}^{(2)}\right]_\mathrm{opt}$ &
$\left[b_1^{(2)}\right]_\mathrm{opt}$ \\ \hline
$\#1$ & $-6.05$ & $-6.05$ & $-4.23$ & $-4.24$ & $2.36$ & $6.29$ & $-8.68$ &
$8.46$ & $-3.94$\\
$\#2$ & $-6.20$ & $5.64$ & $5.91$ & $-5.96$ & $-3.78$ & $-3.63$ & $8.53$ &
$8.65$ & $-3.68$\\
$\#3$ & $-6.28$ & $-6.17$ & $6.31$ & $6.10$ & $2.11$ & $-3.28$ & $-7.49$ &
$9.21$ & $2.56$\\
$\#4$ & $-1.73$ & $-8.76$ & $1.74$ & $9.31$ & $-0.35$ & $4.02$ & $16.96$ &
$-11.51$ & $1.02$\\
$\#5$ & $7.25$ & $5.65$ & $7.16$ & $5.69$ & $-3.25$ & $-8.60$ & $13.61$ &
$-14.04$ & $-6.53$ \\ \hline
\end{tabular}
\end{table*}

Alternative methods to optimize the network have been introduced. One such 
approach is 
zero-temperature Monte Carlo sampling~\cite{whitelam1}, where the parameter 
values are updated randomly. If the random update of parameters leads to a 
lower loss, the 
modified parameter values are kept. Otherwise, they are reset to the values 
prior to the update. Currently among the most popular 
optimization algorithms, the Adam optimizer~\cite{ba1} significantly improves 
convergence dynamics by adaptively changing learning rates for individual 
parameters. We have used these methods to compare with our
results obtained by standard gradient descent for the convergence dynamics. 
Since we concentrate here on the exploration of the geometric properties of 
the state space, optimizing the performance was not in the focus of this 
study, though.

For the estimation of the microcanonical entropy, we employed Monte 
Carlo 
sampling 
methods. Simple sampling, i.e., identifying states by randomly sampling the 
parameter space turned out to be sufficient, but results were confirmed by 
means of a generalized-ensemble importance-sampling method. For this purpose, 
we combined the Wang-Landau algorithm~\cite{wl1,wl2} with multicanonical 
sampling~\cite{muca1,muca2,mb1}.
\section{Optimization Characteristics of the XOR Network and Loss Landscape}
\label{sec:sol}
\subsection{Dependence of Network Performance on the Learning Rate}
\label{ssec:lr}
\begin{figure}
\centerline{\includegraphics[width=8.5cm]{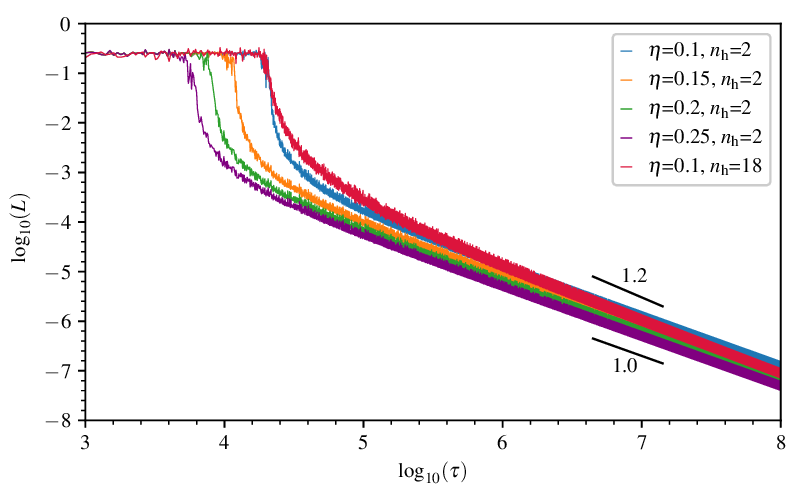}}
\caption{\label{fig:lvstau}%
Convergence of the loss function for the minimal network with 
$n_\mathrm{h}=2$ hidden neurons as a function of epochs $\tau$ for various 
learning rates. For comparison, the loss curve for a larger network with 
$n_\mathrm{h}=18$ neurons in the hidden layer is also included. Reference 
lines with values of the exponent $\gamma$ attached support the power-law 
behavior in the long term.}
\end{figure}
We convinced ourselves by performing multiple long runs that 
the perceptron with a single hidden 
neuron cannot solve the problem. It does not 
have the necessary capacity to store the memory for all cases. The 
non-existence of a solution will eventually be evident from the analysis of 
the microcanonical entropy later. However, the 
addition of another hidden neuron to the neural network as shown in 
Fig.~\ref{fig:ann} ($n_\mathrm{h}=2$) is already sufficient, although 
the optimization process 
still requires a large number of epochs. Figure~\ref{fig:lvstau} shows 
the 
convergence of the loss function for this small network toward zero 
for multiple learning rates $\eta$. For comparison, the loss curve 
for a network with $n_\mathrm{h}=18$ neurons in the hidden layer and learning 
rate $\eta=0.1$ is shown as well.
Weights and biases were initialized at random values, $v_k=0.1r_k$ 
($k=1,\dots,9$), where $r_k\in [0,1)$ is a random number drawn from a uniform 
distribution. The fluctuations are a consequence of the randomly chosen 
members of the batch. If the backpropagation recursions had been 
performed with nonrandomized batches, the curves would be smooth. As 
expected, the choice of the learning rate has a significant influence on the 
optimization dynamics. However, convergence can only be achieved within a 
certain range of values. For this example, in extensive 
tests with up to five million epochs, we found the interval of convergence to 
be $\eta\in[0.002,17.484]$. For the following discussion of the optimization 
characteristics, we employ the results obtained for $\eta=0.1$. Convergence 
sets in after about $70000$ epochs. 
In the initial phase, the loss fluctuates about $0.25$ and it 
seems surprising that it takes 
several ten thousand more epochs to complete the learning process for all 
cases. However, the shape of the loss landscape to be discussed later will 
provide essential clues.

The double-logarithmic plots in Fig.~\ref{fig:lvstau} offer a few 
particularly remarkable insights into the optimization process. Three major 
distinct phases of the optimization process can be identified. First there is 
the initial phase, where the loss fluctuates about $0.25$. This value 
depends on the inital parameter settings. Then there is a sudden drop into 
the attraction basin. In this period, the learning rate has the expected 
impact on the convergence dynamics. Topographically, larger learning rates 
make the optimization algorithm traverse the softly sloped terrain toward the 
basin of attraction faster. Eventually, the optimization process slows down 
significantly. It enters the long-term era governed by the power law
\begin{equation}
L(\tau)\sim \tau^{-\gamma}
\end{equation}
with the exponent $\gamma$. For the smallest network with two hidden neurons, 
$\gamma\approx 1.0$, whereas its value increases to about $1.2$ for the 
largest network studied ($n_\mathrm{h}=18$). Remarkably, $\gamma$ 
does not seem to depend on the learning rate, but only on the number of 
hidden neurons. By comparing the networks with $2$ and $18$ hidden neurons 
at the learning rate $\eta=0.1$, it is noteworthy that the larger network 
converges faster only in the long-term phase because of the larger exponent.

The optimal parameter values found with the above initialization and 
learning rates $\eta=0.1$ for the network with $n_\mathrm{h}=2$ are 
listed in 
Table~\ref{tab:sol} as solution \#1. This particular optimal parameter set 
serves as the basis for the subsequent discussion. The other examples of 
optimal sets of 
parameter values were obtained in zero-temperature Monte Carlo runs. The 
selection in Table~\ref{tab:sol} is not expected to be complete; other 
solutions exist.
In fact, as we will see later, the space of parameter sets with marginal 
loss values close to zero is continuous, i.e., formally there are infinitely 
many parameter sets that effectively solve the problem.
\begin{figure}
\centerline{\includegraphics[width=8.5cm]{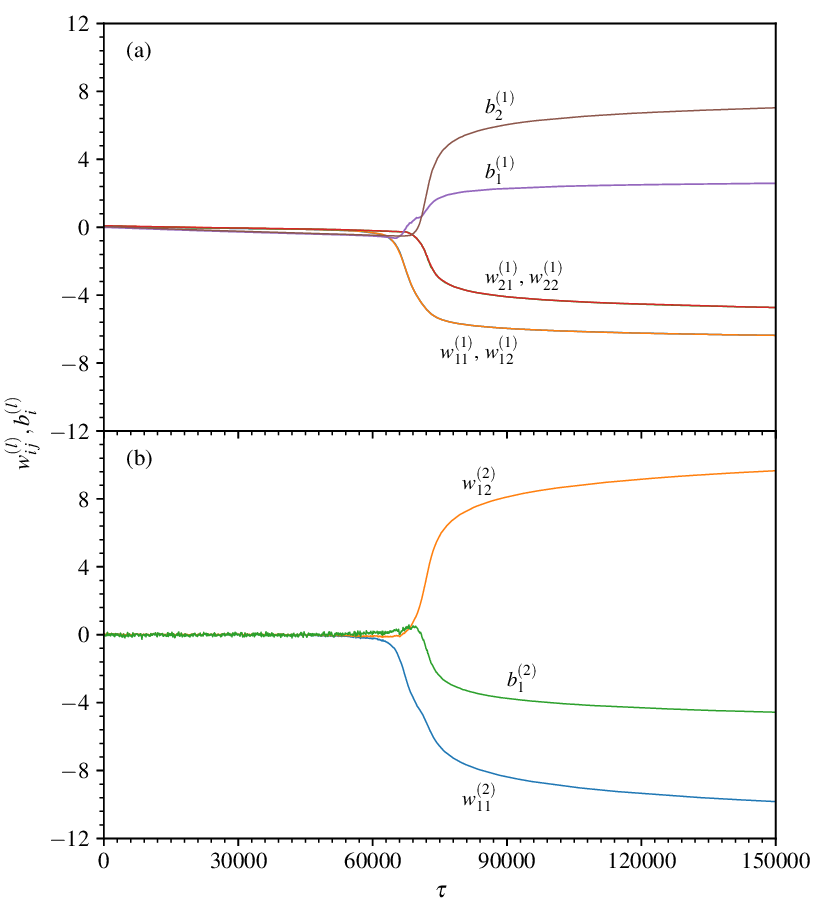}}
\caption{\label{fig:drift}%
Drifts of weights and biases for the (a) hidden and (b) output layer as the 
optimization process progresses through the epochs $\tau$ at learning rate 
$\eta=0.1$. Note that the pairs of weights $w_{11}^{(1)},w_{12}^{(1)}$ and 
$w_{21}^{(1)},w_{22}^{(1)}$, respectively, are indistinguishable for this 
solution (\#1 in Table~\ref{tab:sol}) of the problem.} 
\end{figure}
\subsection{Convergence Dynamics}
\label{ssec:conv}
A particularly striking feature in the optimization process of sigmoid 
neurons in this configuration of the network is that zero loss cannot be 
reached in a finite number of epochs. In fact, as Fig.~\ref{fig:drift} 
shows, the values of weights and biases keep drifting as 
backpropagation proceeds epoch by epoch. It poses the interesting question 
inhowfar the truncation point of the optimization dynamics impacts the 
performance of the network for data the network was not trained for (which 
cannot be verified for the logical problem in this study). From the curves 
for the individual weights and biases shown in Fig.~\ref{fig:drift}, we see 
that, after an extensive initial phase with little change, a sudden 
separation 
sets in at the same time when the loss rapidly drops (see 
Fig.~\ref{fig:lvstau}). First, the weights of hidden neuron 1 start changing, 
which triggers a similar drop in the weight $w_{11}^{(2)}$ connecting this 
hidden neuron with the output neuron. In consequence, also the bias of 
hidden neuron 1, $b_1^{(1)}$, shifts to higher values. Only then, in a second 
step, all other weights and biases respond accordingly, ultimately leading to 
the drop of the loss and its convergence toward zero. However, even though 
this is 
achieved after about $80000$ epochs, the parameters keep changing 
monotonously. 

It is worth noting that the Adam optimizer~\cite{ba1} yields 
qualitatively the same results. The convergence sequence for the 
individual parameters is identical to that shown in 
Fig.~\ref{fig:drift}. However, as expected, Adam is more efficient than 
stochastic gradient descent based on fixed learning rates. Convergence 
sets in much faster.

\begin{figure}
\centerline{\includegraphics[width=8.5cm]{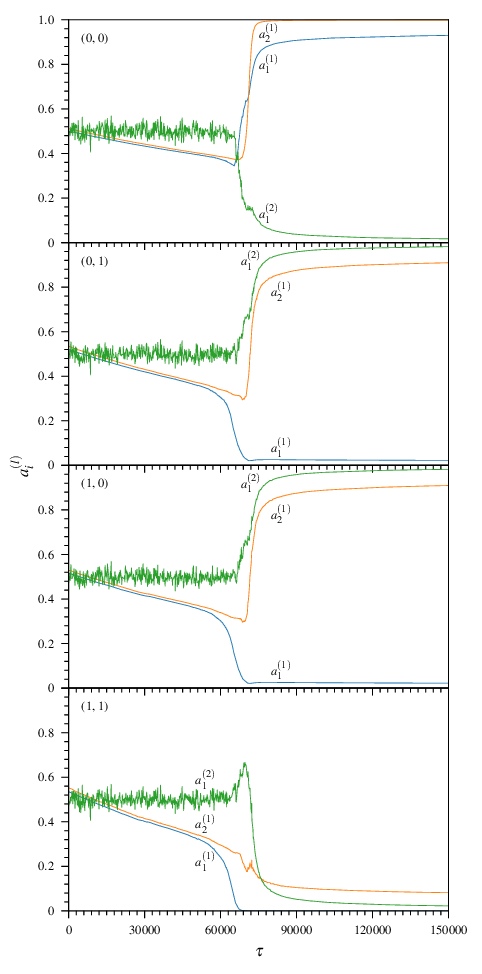}}
\caption{\label{fig:activ}%
Convergence of activations $a_i^{(l)}$ of hidden and output neurons to 
solution \#1 in Table~\ref{tab:sol} for all 
four cases listed in Table~\ref{tab:xor} as functions of epoch $\tau$ for the 
XOR sigmoid network. The learning rate was $\eta=0.1$.} 
\end{figure}

Despite the parameter drifts, the activation of the output, 
$a_1^{(2)}$, reliably converges to 
the expected result in each of the four cases, as is shown in 
Fig.~\ref{fig:activ} for backpropagation at $\eta=0.1$. In early epochs, it 
fluctuates about $0.5$, but only when the activations of the intermediate 
layer, $a_1^{(1)}$ and $a_2^{(1)}$, desync in a symmetry breaking process, 
$a_1^{(2)}$ can ultimately enter its convergence channel. Somewhat 
surprising are the rapid turns of $a_2^{(1)}$ after separating from 
$a_1^{(1)}$ at $\tau\approx 70000$ in the cases (0,0), (0,1), and (1,0), 
which initiate the transition process. The case (1,1) is special insofar as 
the dynamics of $a_1^{(2)}$ embarked into the wrong channel and only after 
an additional fluctuation in $a_2^{(1)}$ it corrects the direction and the 
output value eventually approaches zero. It is the distinct split in the 
respective dynamics of $a_1^{(1)}$ and $a_2^{(1)}$ that make this neural 
network find a solution to the XOR problem; a single intermediate neuron as 
in a perceptron representation would not have been sufficient to drive the 
behavior of the output neuron correctly.

It should be noted that the similar activation values in all cases at about 
$0.5$ in the first phase are a consequence of the small initial values of the 
parameters randomly chosen in the interval $(0,0.1)$; the argument of the 
sigmoid function at the output 
is close to $0$ and $\sigma(0)=0.5$. The advantage of this initialization is 
that convergence dynamics is well-behaved. 

\begin{figure}
\centerline{\includegraphics[width=8.5cm]{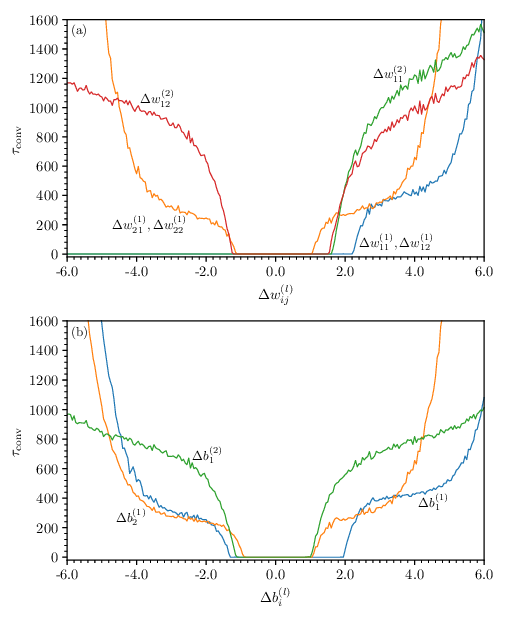}}
\caption{\label{fig:eofc}%
Epoch of convergence $\tau_\mathrm{conv}$ plotted as function of the 
deviation of initial (a) weights $w_{ij}^{(l)}$ and (b) biases $b_i^{(l)}$ 
from the respective optimal values for solution \#1 given in 
Table~\ref{tab:sol}. In each case, all other weights and biases are 
initialized at their optimal values.} 
\end{figure}

In order to better understand the approach to the solution, let us 
have a 
look at the convergences dynamics if one parameter is shifted away from its 
optimal value, whereas the other eight are kept constant at their optimal 
values. This 
is done for each weight and bias separately. The results are shown in 
Fig.~\ref{fig:eofc}. Epoch of convergence $\tau_\mathrm{conv}$ is defined as 
the number of epochs it takes for the network to converge to solution \#1 in 
Tab.~\ref{tab:sol} if one specific parameter is initialized at 
$\left[w_{ij}^{(l)}\right]_\mathrm{init}=
\left[w_{ij}^{(l)}\right]_\mathrm{opt}+\Delta w_{ij}^{(l)}$ (weights) or
$\left[b_i^{(l)}\right]_\mathrm{init}=
\left[b_i^{(l)}\right]_\mathrm{opt}+\Delta b_i^{(l)}$ (biases), 
respectively. Convergence is achieved once the loss, averaged over up to 
$100$ backpropagation runs with the same parameter 
initializations (different seeds of the random number generator generate 
different random batches, though), falls below the threshold value 
$0.005$, and a test with the nonrandomized batch also yields a loss value 
below this 
threshold. This ensures that the network has indeed converged. 

The results are quite remarkable and give us first clues regarding prominent 
features of the loss landscape. First of all, we observe that all curves 
possess a plateau in the vicinity of their optimal values (i.e., around 
zero for the shifted weights and biases). Whereas fast convergence is 
expected very close to the zero point, the rather large extension of the 
plateau for positive and negative deviations from the optimal value is rather 
surprising. Imagining a higher-dimensional loss landscape, this would mean 
that the optimal solutions are embedded in a region with almost zero loss 
value. 
We can also see that the more the selected initial weight or bias is shifted 
away from its optimal value the more rapidly the convergence dynamics changes.
There 
are interesting exceptions, though. Below certain threshold values, the 
variation of the initial settings of the weights $w_{11}^{(1)}$, 
$w_{12}^{(1)}$, and $w_{11}^{(2)}$ do not affect the solution at all. From 
Fig.~\ref{fig:eofc}(a), we find that the convergence dynamics of the network 
is not altered if 
$\Delta w_{11}^{(1)}<2.2$, $\Delta w_{12}^{(1)}<2.2$, or 
$\Delta w_{11}^{(2)}<1.6$ (provided in each case all other parameters are 
initialized at their optimal values). This supports the conclusion that there 
is indeed a continuous space of solutions, forming a valley or channel 
in the loss landscape. 

Furthermore, it can be seen that for some of the parameter shifts 
convergence is slower, but remains fairly constant upon alterations of the 
shifts away from the optimal values. This hints at plateaus at higher 
loss values in the loss landscape. 
\subsection{Analysis of the Loss Landscape}
\label{ssec:loss}
In the following, we investigate the features of cross-sections through the 
loss landscape. For this purpose, we compare the results 
obtained by using full nonrandomized and randomized batches for the 
calculation of the loss in forward passes.
\subsubsection{Nonrandomized Batch}
\label{ssec:nonrand}
\begin{figure*}
\centerline{\includegraphics[width=16.8cm]{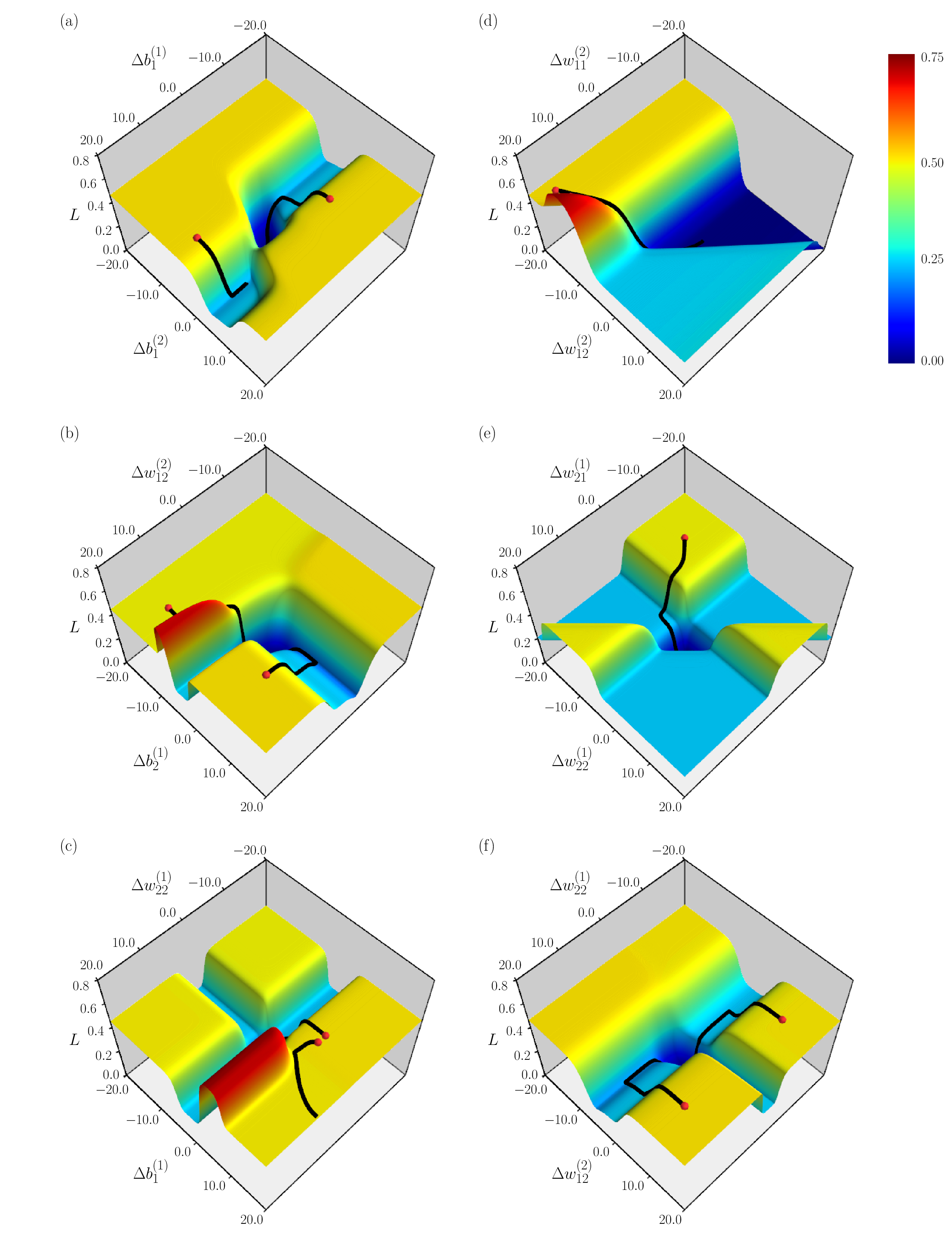}}
\caption{\label{fig:6in1}%
Selected cross-sections of the loss landscape through the nine-dimensional 
parameter space. Seven parameters are kept constant at their optimal 
values, and two are varied about their 
respective optima in the ranges shown. Various 
trajectories are included to visualize the backpropagation dynamics starting 
from the locations marked by the red dots.} 
\end{figure*}
The loss was defined in Eq.~(\ref{eq:loss}) as an average of the square 
deviations of the actual from the expected output values over the batch of 
all input cases for any given set of parameter values. Therefore, the loss 
function depends on all parameters and can be interpreted as a landscape 
in this embedding space. For the network we chose, the 
parameter space is nine-dimensional. In order to visualize some of its 
features, we have calculated all 36 cross-sections in which 7 of the 9 
parameters are kept constant at their optimal values according to solution 
\#1 in Table~\ref{tab:sol}, whereas two are varied about their optimal values 
in 
a rather large range. Cross-sections for six different choices of 
variable parameter pairs are shown in Fig.~\ref{fig:6in1}. Whereas all 36 
cross-sections significantly differ from each other, the selection already 
contains the most distinct features. It should also be noted that, not 
surprisingly, landscapes look different for other solutions 
listed in Table~\ref{tab:sol}, but the cross-sections possess similar main 
features like the ones shown in Fig.~\ref{fig:6in1}. Also included in the 
figures are exemplified backpropagation or convergence trajectories with 
randomly chosen starting points. Like for the loss landscape, the complete 
batch of the four cases listed in Table~\ref{tab:xor} without randomization 
was used for the calculation of the trajectories 
and the steepest-descent method was only applied to the two variable 
parameters, whereas the other seven were kept strictly constant at their 
optimal values.

The most striking features in all figures are the channels, plateaus, 
and isolated rims and wells at loss values between $0$ 
and $1$ at $0.25$ increments, separated by steep slopes. The prominent 
loss values 
of these features can be readily explained: According to our definition of 
the loss as the sum of square deviations over the full batch as used in the 
construction of the landscapes, zero loss can only be achieved if all outputs 
calculated by the network coincide exactly with their expected values in 
all cases. In all cross-sections shown in Fig.~\ref{fig:6in1}, this point is 
located at $(0,0)$, and regions with 
values close to zero are colored dark blue. An 
isolated well surrounding this point is present in all cross-sections shown, 
except in Fig.~\ref{fig:6in1}(d), where the zero-loss point is at the corner 
of a shallow triangular, almost plateau-like, region. In 
Fig.~\ref{fig:6in1}(b), 
there appears to be a deep trench (also dark blue) with very small loss 
values converging to the zero-loss point. Such a trench, albeit at higher 
loss values, is also visible in Fig.~\ref{fig:6in1}(f). 

Loss values $\sim 0.25$ 
can typically be attributed to situations, where the XOR logic of one case in 
the batch is determined wrongly (the output is either $1$ instead of $0$, or 
vice versa). Those regions in the landscapes are shaded turquoise. It is 
noticeable that these regions often form straight channels, but also 
plateaus.

If among the four cases in the batch two outputs are correct and two wrong, 
the loss value is $0.5$. This is the ``highest-entropy'' situation as there 
are $6$ possible combinations of pairs of cases in the batch that lead to 
$L=0.5$. Consequently, there are significantly more parameter combinations 
representing this scenario than any other, and therefore the yellow plateaus 
associated with loss values around $0.5$ dominate in all figures. In contrast, 
larger loss values are clearly suppressed in the parameter ranges plotted. 
Only in Figs.~\ref{fig:6in1}(b)--~\ref{fig:6in1}(d), we see red-shaded rims 
with values at about $L=0.75$, in which situation three cases in the batch are 
evaluated wrongly. Areas with even higher loss values are not present in the 
parameter spaces covered in any of the cross-sections analyzed in the 
vicinity 
of the optimal parameter settings. It should be noted that all real loss 
values between $0$ and $1$ are permitted as the sigmoid output is continuous, 
but near the optimal solution, only these distinct, almost discrete, shapes 
of the cross-sections through the loss landscape are observed. However, for a 
problem with more than four cases in the full batch, the landscape features 
are certainly more diverse. Note that thanks to the sigmoid activations, all 
plateaus, channels, and trenches have slight slopes, i.e., there is always a 
gradient toward lower loss values leading to a local minimum at loss values 
larger than zero or a global minimum at zero loss. 
\begin{figure*}
\centerline{\includegraphics[width=16.8cm]{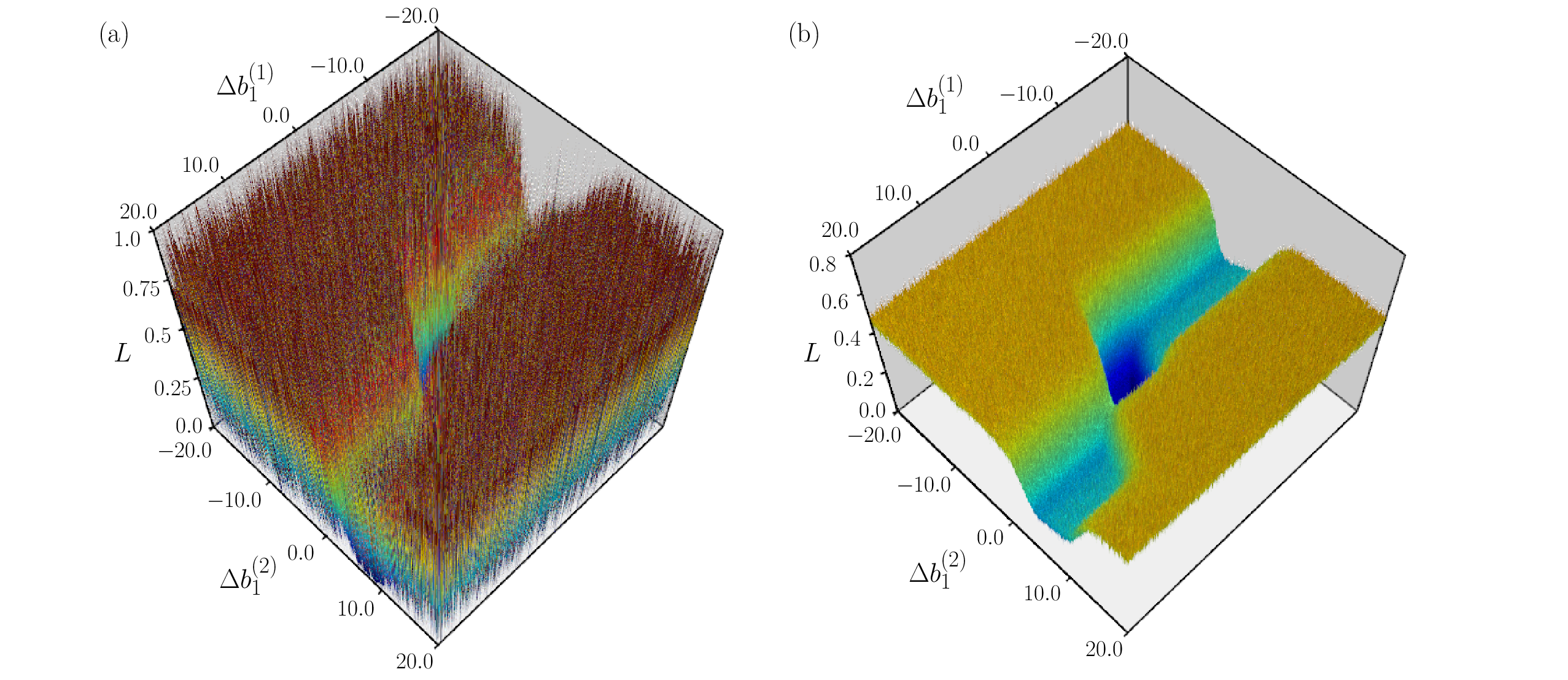}}
\caption{\label{fig:rnd}%
Cross-section through the loss landscape with $b_1^{(1)}$ and $b_1^{(2)}$ 
variable for randomized batches: (a) 
single run, (b) average over 100 runs. Compare with the result obtained for 
the 
nonrandomized batch in Fig.~\ref{fig:6in1}(a).} 
\end{figure*}

Taking a closer look at selected convergence trajectories is quite 
revealing. These paths are inserted into the plots shown 
in Fig.~\ref{fig:6in1}. Since in this consideration the complete 
nonrandomized batch is used and 
not a stochastic one, the trajectories are smooth. The locations marked 
in red define the starting points or initial conditions for the dynamics. The 
learning rate was set to $\eta=0.1$ in most cases (larger values were used 
where the dynamics was too slow). The simulation was stopped once 
convergence to zero loss was achieved or preset limits of up to $60$ million 
epochs were reached otherwise. 

Figure~\ref{fig:6in1}(a) contains two different
trajectories. The path beginning at 
$(\Delta b_1^{(1)},\Delta b_1^{(2)})=(-7.4,8.9)$ with loss value $L\approx 
0.5$ rapidly reaches the cliff and drops to the lower level at about 
$L\approx 0.25$. After residing there for a number of epochs it 
quickly descends to the bottom and converges to the zero-loss point in the 
well. This is the common scenario of convergence to the optimal solution. 

The behavior is completely different for the path initiated at 
$(\Delta b_1^{(1)},\Delta b_1^{(2)})=(15.0,-2.0)$ (the loss value 
is $L\approx 0.5$), which represents an unsuccessful attempt. It also rapidly 
drops to a trench at $L\approx 0.25$, but any further convergence is clearly 
prevented by a barrier separating this channel from the well. The path gets 
stuck at a local minimum. It is somewhat surprising that the loss landscape of 
this simple network we study here exhibits such features of complexity and is 
a reminder that optimization is not a straightforward process. 

In Fig.~\ref{fig:6in1}(b), the trajectory initialized at 
$(\Delta w_{12}^{(2)},\Delta b_2^{(1)})=(12.5,-12.5)$, for which $L\approx 
0.48$, contrasts the behavior described in the 
previous case. The origin of the path is deliberately chosen to be located 
behind a barrier, but the gradients are favorable and enable the ``walker'' to 
pass by the ridge and directly drop into the basin of convergence to zero 
loss.

Another very interesting scenario is shown in Fig.~\ref{fig:6in1}(c). Two 
trajectories are initialized in close proximity from each other, $(\Delta 
w_{22}^{(1)},\Delta b_1^{(1)})=(-5.0,10.0)$ vs.\ $(-3.0,10.0)$, both at 
loss values very close to $L=0.5$. Whereas the former rapidly 
converges to 
the optimum point, the other takes a turn, hits a barrier and is diverted 
away from the zero-loss attraction basin. Millions of epochs later it still 
had not converged to this or any other solution. It seems the 
trajectory got trapped in a local minimum at a loss values just below 
$L\approx 0.5$. This divergence in the dynamics of 
backpropagation for this system may hint at potentially chaotic behavior in 
larger networks as landscape shapes are more fragmented. Note that there 
is no stochastic component in the way we backpropagate here. At 
any given location, trajectories follow the steepest decent and cannot climb 
uphill.
\subsubsection{Randomized Batches}
\label{ssec:rand}
Randomized batches are typically used for the learning process of large 
neural networks, where working with the full batch is not economical. It is a 
necessity if mini-batches are used. In this section, we again use the full 
batch, but with randomized cases, to determine cross-sections through the 
loss landscape. Since the results strongly fluctuate due to the stochastic 
sampling of the batch compositions, the loss landscape looks fuzzy for a 
single scan. This is shown in Fig.~\ref{fig:rnd}(a) for the 
cross-section, where $b_1^{(1)}$ and $b_1^{(2)}$ are variable and all other 
parameters are kept constant at their optimal values, according to solution 
\#1 in Table~\ref{tab:sol}. Compare this to the smooth 
cross-section plotted in Fig.~\ref{fig:6in1}(a), which was 
obtained using the nonrandomized batch. It is interesting that, despite the 
strong fluctuations, the main landscape features are already visible. After 
averaging over 100 different scans [see Fig.~\ref{fig:rnd}(b)], we find that 
the 
landscape shape indeed converges to the result we obtained for the 
nonrandomized batch. This is reassuring as it confirms that the 
randomization of batches is a viable option and does not alter the basic 
landscape features including the attraction basin toward the optimal solution.
\subsection{Density of Loss and Microcanonical Entropy}
\label{ssec:dos}
In modern statistical physics, the density of states $g(E)$ is often used 
for the identification of significant thermodynamic features such as phase 
transitions in complex physical systems. If $\mathbf{X}$ is the phase space 
vector of all degrees of freedom (such as coordinates and momenta in 
Hamiltonian systems), it is defined as
\begin{equation}
g(E) = \int {\cal D}\mathbf{X}\,\delta(H(\mathbf{X})-E),
\end{equation}
where $H$ is the energy function or Hamiltonian assigning the state 
$\mathbf{X}$ an energy value, and ${\cal D}\mathbf{X}$ symbolizes the 
multi-dimensional integral measure in the state space. The integral over the 
Dirac $\delta$ distribution therefore counts how many states share the energy 
$E$. Since $g(E)$ addresses all states available in the phase space of the 
system, it can be interpreted as the phase space volume per energy unit and 
therefore allows for the introduction of the so-called microcanonical or 
restricted entropy according to Boltzmann's formula:
\begin{equation}
\label{eq:ent}
S(E)=k_\mathrm{B}\ln g(E),
\end{equation}
where $k_\mathrm{B}$ is the Boltzmann constant. This relation is the 
foundation of statistical physics and connects it to thermodynamics. 
As the system behavior is governed by the competition of enhancing 
entropy and reducing energy, which for most systems leads to stable 
thermodynamic equilibrium phases, Eq.~(\ref{eq:ent}) contains all information 
about the phase transitions a system may experience. Since the inverse 
temperature is defined as 
$\beta(E)=dS/dE$, curvature properties of $S(E)$ even allow for the 
identification of transition temperatures. The recently introduced 
generalized microcanonical inflection-point analysis method~\cite{qb1} 
enables the systematic identification and classification of all phase 
transitions. It can even be used for finite systems, where the hypothetical 
thermodynamic limit needed in the conventional theory of phase transitions 
cannot be employed.

In this context, we may consider a neural network a complex physical 
system and interpret the learning process as a cascade of phase transitions. 
This can even be done for the simple artificial neural network discussed in 
this paper. We replace the energy by the loss and redefine $g(L)$ as the 
density of loss. Then $S(L)=\ln g(L)$ is a dimensionless 
microcanonical entropy for this 
problem. The shape of this curve should encode the information about the 
activation steps needed in the learning process to overcome the transition 
barriers (if there are any) toward the accomplishment of the learning 
objective. In this sense, learning is an annealing process passing through 
different phases from high-entropy states (high temperature) to the 
optimal solution, which represents a low-energy state (as 
temperature approaches zero).

Another advantage of this approach is that it covers the entire 
state space, but $S$ is only a function of a single variable, $L$. This 
reduces the effort of trying to find features of the loss landscape by means 
of cross-sections, as done in the previous sections, dimensional reduction or 
the introduction of specific order parameters.

\begin{figure}
\centerline{\includegraphics[width=8.5cm]{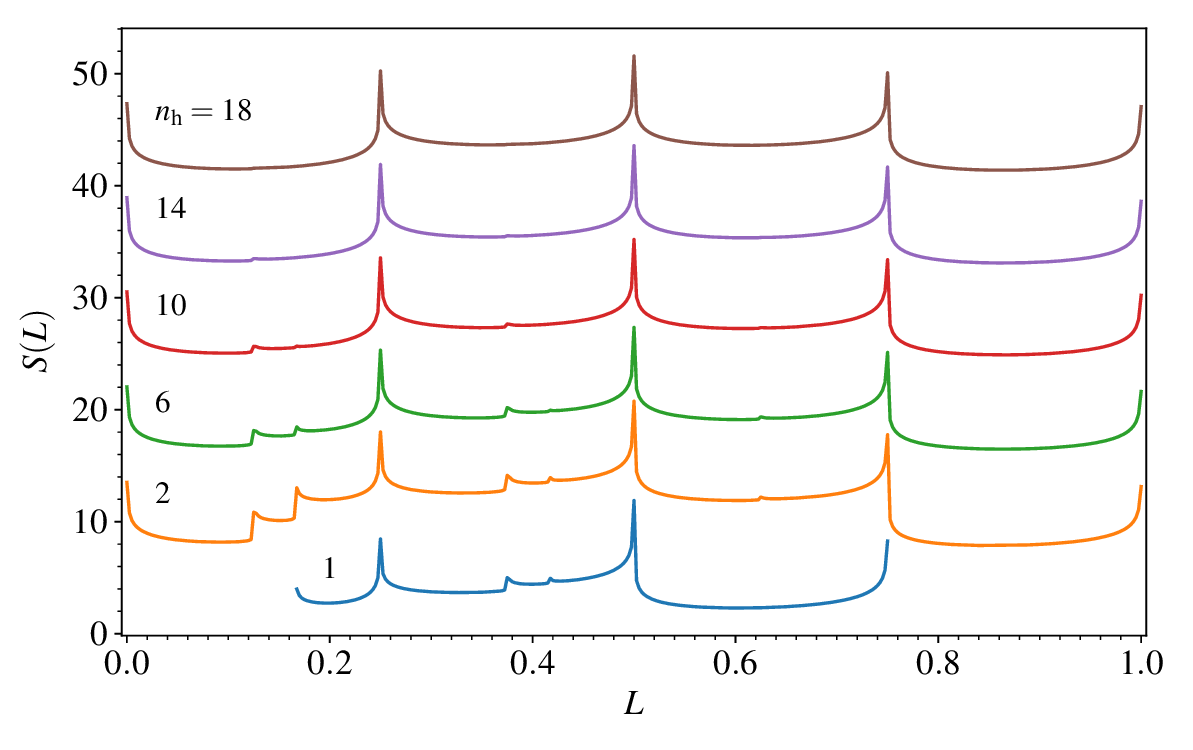}}
\caption{\label{fig:entropy}%
Microcanonical entropies $S(L)$ for XOR networks with different 
numbers of neurons in the hidden layer, $n_\mathrm{h}$. The curves have been 
shifted relative to each other vertically for better visibility.}
\end{figure}

Microcanonical entropies for networks with $n_\mathrm{h}$ neurons in 
the 
hidden layer, obtained by generalized-ensemble 
Monte Carlo simulations~\cite{wl1,wl2,muca1,muca2}, are shown in 
Fig.~\ref{fig:entropy}. 
From the curve for $n_\mathrm{h}=1$ it is obvious why the perceptron 
fails solving the XOR problem: There are simply no zero-loss states in the 
parameter space. For networks with two and more hidden neurons, 
multiple sets of parameter values for zero loss can be found. The 
entropy is largest around 
$L=0.5$ and noticeable peaks are also found at $L=0,0.25,0.75$, and $1.0$. 
This is expected 
from the 
discussion of the cross-sections through the loss landscape in the previous 
sections.
Despite the 
expectation of an enhanced entropy at these specific values from the 
discussion of the loss landscape, the very strong suppression of loss values 
in-between is, however, astonishing and responsible for the general 
complexity of the learning 
process. 
The entropy curves for networks with smaller hidden layer show additional 
minor peaks. These correspond to intermediate plateaus in the loss landscape 
and effectively slow down the optimization process. Even though adding 
neurons to the hidden layer significantly increases the dimension of the 
parameter space, these extra dimensions enable the optimizer to bypass these 
entropic barriers. Therefore, larger networks can improve the learning 
efficiency. As Fig.~\ref{fig:entropy} clearly shows, these entropic 
``finite-size'' effects gradually disappear as the number of hidden neurons 
is increased. This supports a virtually barrier-free learning 
process in larger networks~\cite{hamprecht1}. This can be essential for 
large-scale problems, where the optimization usually starts from larger loss 
values. However, for the small, discrete problem studied here, no obvious 
benefits from the smoothing of the entopy curves for the larger networks were 
observed in this phase of the learning process (cf.\ comparison for 
$n_\mathrm{h}=2$ and $18$ at $\eta=0.1$ in Fig.~\ref{fig:lvstau}).

In the context of thermodynamic phase 
transitions, convex regions of the microcanonical entropy indicate 
first-order phase 
transitions~\cite{gross1,mb1,qb1}. In thermodynamic systems, first-order 
transitions are associated with latent heat, i.e., additional energy needed 
to overcome attractive interactions that keep particles bound to each other. 
In this dissolution process, the temperature remains unchanged, thereby 
slowing down any annealing process. It does not seem far fetched to adopt 
this interpretation here and make the convex entropy regions responsible for 
the slowing-down of the learning processes in neural networks, artificial and 
real. It will be an intriguing future task to investigate larger networks 
(large input data sets or number of cases and therefore possessing a 
network state space of much higher dimensionality) in 
this respect. The expectation is that the peaks and thus also 
the negative-slope sections in the convex regions of the entropy curve 
deteriorate. Negative slopes correspond to negative microcanonical 
temperatures and represent unstable non-equilibrium behavior in physical 
systems. Entropy curves for larger networks should therefore look smoother and 
their curvatures may encode inflection points in higher-order derivatives 
indicating transitions higher than first order. The reason for these 
assumptions is that an artificial neural network can serve as a model for a 
real neural network. Therefore, statistical properties of larger artificial 
neural networks are expected to closely resemble features of complex physical 
systems.
\section{Summary}
\label{sec:sum}
Understanding the dynamics of optimization processes for artificial neural 
networks and identifying structural properties in the loss landscape 
that either accelerate or decelerate the training process should be 
considered a key task in helping develop fields of machine 
learning and artificial intelligence more systematically. In this paper, we 
have studied aspects associated with this problem for a very simple sigmoidal 
neural network that represents the XOR logic with a two-bit input and single 
output. An analysis of the convergence dynamics showed that for fixed learning 
rate, network parameters need to correlate cooperatively 
in order to 
correctly activate neurons. We also found that, whereas convergence toward the 
solution (i.e., zero loss) is achieved, weights and biases keep drifting and, 
by themselves, do not converge at all, which was unexpected. 
This behavior can be explained by the long-term behavior of the 
optimization process, which exhibits a power-law decay with a characteristic 
exponent, which is independent of the learning rate, but depends on the 
number of hidden neurons. Ultimately, 
these network properties result in a loss landscape with distinct properties.

Investigating cross-sections through this landscape, we noticed that in the 
vicinity of the optimum, characteristic features like wells, channels, 
trenches, barriers, plateaus, and rims dominate. These geometric shapes 
significantly impact the convergence dynamics of the backpropagation 
optimization method we employed to find optimal solutions. The study as to 
inhowfar these features connect to each other in the full high-dimensional 
parameter space is left to future work. Also, the shape of the loss landscape 
depends on other factors as well, e.g., the choice of the activation 
function. Whereas simpler activation functions than the sigmoid function used 
in this study may reduce computational cost in the forward pass, the 
optimization could be negatively affected by other types of landscape 
barriers. For example, our tests with ReLU activation have shown that 
successful convergence much more sensitively depends on the initial parameter 
settings and learning rates than it was in the case of sigmoid activation.
We also investigated potential impacts of nonrandomized and 
randomized full batches of input data on the landscape features and found 
that the landscape features obtained by averages over randomized batches 
resemble those obtained from the forward passes of the nonrandomized batch. 
Randomization, which is a necessity when using 
mini-batches for large data sets, may even accelerate the annealing process 
in the search for the optimal solution.

Eventually, we introduced the microcanonical entropy of neural 
networks as 
the logarithm of the density of loss as a variant of the
density of states that commonly aids the understanding of the statistical 
physics of phase transitions in thermodynamic systems. Comparing entropy 
curves for networks with different numbers of hidden neurons reveals the 
disappearance of smaller entropic barriers, which indeed supports earlier 
claims that optimization processes in neural networks are essentially 
barrier-free~\cite{hamprecht1}. The analysis of curvature features of 
microcanonical entropy curves has turned out to be very useful in the 
identification and classification of phase transitions in physical 
systems~\cite{qb1}. Transferring this idea to neural networks, 
different stages of learning success represent stable phases. Rapid, 
low-entropy drops in the loss landscape can then be understood as phase 
transitions accompanying the learning process. Approaching zero loss is 
equivalent to cooling down a macroscopic system, i.e., entering more and more 
ordered phases with lower entropy until, ultimately, the ground state is 
reached. The results obtained in this simple case study are already 
intriguing and may encourage more generalized analyses for larger networks. 
The identification of virtual, effective parameters for the representation of 
the loss landscape may help guide network training processes more 
efficiently.
\acknowledgments
KA thanks the Center for Simulational Physics at UGA for hospitality 
and support.

\end{document}